\documentclass[useAMS,usenatbib]{mn2e}
\usepackage{graphicx}
\usepackage{natbib}

\def\aap{A\&A}
\def\apj{ApJ}

\def\mnras{MNRAS}

\def\etal{et al.}
\def\nat{{\em Nature}}
\def\s{Section}

\renewcommand{\vec}[1]{\mbox{\boldmath$#1$}}

\newcommand{\D}{\displaystyle}
\newcommand{\DF}[2]{\frac{\D#1}{\D#2}}
\newcommand{\Bracket}[1]{{\left({#1}\right)}}

\usepackage[dvips]{color}

\begin{document}

\title[Emission from warped discs]
{Reprocessed emission from warped accretion discs induced by the 
Bardeen-Petterson effect}
\author[Wu, Chen and Yuan]
{Sheng-Miao Wu$^{1}$\thanks{E-mail:smwu@shao.ac.cn}, Lei Chen$^{1,2}$
\thanks{E-mail:lchen@shao.ac.cn, lchen@mpifr-bonn.mpg.de} and  
Feng Yuan$^{1}$\thanks{E-mail:fyuan@shao.ac.cn}\\
$^{1}$Key Laboratory for Research in Galaxies and Cosmology, Shanghai 
Astronomical Observatory, Chinese Academy of Sciences, \\ 80 Nandan Road, 
Shanghai 200030, China\\
$^{2}$Max-Planck-Institut fuer Radioastronomie, Auf dem Huegel 69, 53121 
Bonn, Germany}

\maketitle

\date{Accepted . Received ; in original form}

\markboth{Wu et al.: Emission from warped discs}{}

\begin{abstract}
The broad Balmer emission-line profiles resulting from the
reprocessing of UV/X-ray radiation from a warped accretion disc
induced by the Bardeen-Petterson effect are studied.
We adopt a thin warped disc geometry and a central ring-like
illuminating source in our model. We compute the steady-state shape
of the warped disc numerically, and then use it in the calculation 
of the line profile. We find that, from the outer radius to the
inner radius of the disc, the warp is twisted by an angle 
$\sim\pi$ before being flattened efficiently into the equatorial
plane. The profiles obtained depend weakly on the illuminating source 
radius in the range from $3r_{g}$ to $10r_g$, but depend strongly on 
this radius when it approaches the marginally stable orbit of an 
extreme Kerr black hole. Double- or triplet-peaked line profiles are 
present in most cases when the illuminating source radius is low. The
triplet-peaked line profiles observed from the Sloan Digital Sky
Survey may be a {``}signature" of a warped disc.
\end{abstract}

\begin{keywords}
{accretion, accretion discs --- black hole physics --- galaxies: active
--- line: profiles}
\end{keywords}

\section{Introduction}
\label{intro}

A small fraction of active galactic nuclei~(AGNs) with broad lines
show double-peaked emission-line profiles, which suggests an accretion
disc origin \citep*{era94,era03,str03,gez07}. In a stationary
circular relativistic disc model, Doppler boosting will make the
blue peak of the profile higher than the red one. However, observations 
of the variability of line emission in some objects show
that at least at some epochs the profile asymmetry is reversed, with
the red peak higher than the blue one, contrary to the predictions 
of homogeneous, circular relativistic disc models
\citep{mil90,str03,gez07}. To adjust the theoretical to the observed
profiles, non-axisymmetric or warped discs are usually required.
Various physically plausible processes in the accretion flow have been 
considered, such as spiral shocks, eccentric discs, bipolar outflows 
or a binary black hole system
\citep{cha94,era95,sto97,zhe90,vei91,beg80, gas96,zha07b}, as well as
a warped accretion disc \citep*{wu08}. This warped disc makes it
possible for the radiation from the inner disc to reach the outer
parts, enhancing the reprocessing emission lines. Thus, warping
can both provide the asymmetry required for the variations of the
emission lines and naturally solve the long-standing energy-budget
problem, because the subtending angle of the outer disc portion to
the inner one is increased by the warp.

There are strong observational and theoretical grounds for believing 
that accretion discs around black holes may be warped and twisted. 
Evidence for the existence of warped discs in astrophysical systems 
has been found from observations \citep{her05,cap06,cap07,mar08}. 
From a theoretical point of view, four main mechanisms for 
exciting/maintaining warping in accretion discs have been proposed, 
namely warping that is tidally induced by a companion in a binary 
system \citep{teb93,teb96,lar96}, radiation-driven or self-induced 
warping \citep{pri96,pri97,mal96,mal97,mal98}, magnetically driven 
warping \citep{lai99,lai03,pfl04}, or warping driven by frame dragging 
\citep{bap75,kum85,arm99}. Note also that warps generated by 
gravitational interactions have been investigated in the literature. 
In the galactic context, \citet{hun69} studied the linear bending waves 
of a self-gravitating, isolated, thin disc. On nuclear disc scales, 
\citet{pap98} studied the evolution of a thin self-gravitating viscous 
disc interacting with a massive object orbiting the central black hole. 
\citet{ulu09} showed that highly warped discs near black holes can 
persist for a long time without any persistent forcing other than by 
their self-gravity. As a result, the accretion disc in some AGNs may 
be non-planar.

To solve the energy budget for the emission-line region for the disc 
model, reprocessing of the ultraviolet(UV)/X-ray continuum from the 
inner accretion disc is required. However, only a very small fraction 
of radiation from the inner disc is expected to intercept the outer 
part of the disc in the case of a flat geometrically thin disc. Two 
types of processes have been proposed to increase the fraction of light 
incident on the disc emission-line region. \citet*{che89} proposed that 
the inner part of the accretion disc is hot and becomes geometrically 
thick as a result of insufficient radiative cooling, and the X-ray 
emission from such an ion-supported torus is responsible for such energy 
input. However, double-peak emitters do not always have a low accretion 
rate \citep{zha07a,bian07}. In contrast, \citet{cao06} proposed 
that a slow-moving jet can scatter a substantial fraction of UV/X-ray 
light from the inner accretion disc back to the outer part of the disc. 
Although this may be likely for radio-loud double-peaked emitters, 
the majority of double-peaked emitters are radio-quiet. In this work, 
we consider a system with an accretion disc around a Kerr black hole. 
The black hole spin is misaligned with the outer parts of the disc, as 
there is no physical reason to suppose that the angular momentum of the 
accreting mass and the angular momentum of the spinning black hole are 
always aligned. Frame dragging produced by a Kerr black hole causes the 
precession of the orbit of a particle if its orbital plane is inclined 
with respect to the equatorial plane of the black hole. This is known as 
the Lense-Thirring effect. The combined action of the Lense-Thirring
effect, which tends to twist up the disc, and the internal viscosity
of the accretion disc, which tends to smooth it out, forces the
alignment between the angular momenta of the Kerr black hole and the
accretion disc. This is known as the Bardeen-Petterson effect. This
effect tends to affect the inner regions of the disc owing to the
short range of the Lense-Thirring effect($\propto r^{-3}$), whereas 
the outer part of the disc tends to keep its original inclination.
The transition radius between these two regimes is known as the
Bardeen-Petterson radius $R_{\rm BP}$~(or warp radius $R_{\rm w}$),
and its exact location depends mainly on the physical properties of
the accretion disc \citep{bap75,sch96,nel00,fra05a}. Basically, the
Bardeen-Petterson radius is determined by comparing the time-scale
related to the Lense-Thirring precession with that of warp propagation
through the disc. We consider here the propagation of warps in thin 
Keplerian discs in the case for which the disc is sufficiently viscous 
that the warp propagates in a diffusive manner. The twisted disc itself 
is treated using a non-relativistic approach with an additional term 
describing the gravitomagnetic precession of the disc rings. Both analytic 
and numerical steady-state solutions have shown that the evolution of a 
misaligned disc arising from the Bardeen-Petterson effect usually produces
an inner flat disc and a warped transition region with a smooth
gradient in the tilt and twist angles \citep{sch96,nat99,mar07}.
If the disc is thick and/or its viscosity is low, the warp propagates in
a wave-like rather than a diffusive manner \citep*{nel00,lub02}. In this 
case, the twisted disc is not necessarily aligned with the equatorial 
plane at small scales. The inclination angle of a low-viscosity disc 
may even oscillate or counteralign close to the black hole 
\citep{lub02,kin05,lod06,lod07}. 

The effect of a Bardeen-Petterson disc on iron line profiles treated 
relativistically has been examined by \citet{fra05b}, and \citet{bac99} 
used a non-relativistic treatment to study the broad-line H$\beta$ 
profiles from a warped disc. A relativistic treatment of the effect of 
a warped disc with a parametrized geometry on the Balmer lines that can 
be compared with observations was presented by \citet{wu08}.
Following on from these calculations, the disc warping induced by the 
Bardeen-Petterson effect and its influence on the broad Balmer 
emission owing to the reprocessing of the central high-energy
radiation is investigated. In \s~\ref{method} we summarize the
assumptions behind our model and present the basic equations relevant 
to our problem. We present our results in \s~\ref{result}, and
the conclusions and a discussion in \s~\ref{sum}.

\section{Assumptions and Method of Calculation}
\label{method} In this paper, we focus on how a warped disc resulting 
from the Bardeen-Petterson effect affects optical emission-line profiles, 
such as H$\alpha$ and H$\beta$. A geometrically thin disc and a 
central ring-like illuminating source~(radius denoted by $R_{r}$) around 
a black hole are assumed. A short review of the basic equations relevant 
to our problem is given below, and readers are referred to \citet{wu08} 
for a description of our numerical scheme in more detail.

\subsection{Geometrical considerations}

The disc can be treated as being composed of a series of concentric
rings of width ${\rm d}R$ and mass $2\pi \Sigma R {\rm d}R$ at radius 
$R$ from the central point mass $M$ with surface density $\Sigma(R,t)$ 
at time $t$ and with angular momentum $\vec{L}=(GMR)^{1/2}\Sigma
\vec{l}=L\vec{l}$, and lying in different planes. The rings interact 
with each other through viscous stresses. Each ring is
defined by two Eulerian angles, $\beta(R,t)$ and $\gamma(R,t)$, at
radius $R$. At each radius $R$ from the center, the disc has a unit
tilt vector $\vec{l}(R,t)$ that varies continuously with radius $R$
and time $t$. Following \citet{pri96}, the vector $\vec{l}(R,t)$ is
given by
\begin{equation}
  \vec{l}=(\cos\gamma\sin\beta,\sin\gamma\sin\beta,\cos\beta).
  \label{betagamma}
\end{equation}
We define the normalized vector towards the observer as
\begin{equation}
  \vec{i}_{obs}=(\sin i,0,\cos i),
\end{equation}
where $i$ is the angle between the line of sight, and the normal to
the equatorial plane lies in the $\rm XZ$ plane.

We define the coordinates on the surface of the disc as ($R,\phi$)
with respect to a fixed Cartesian coordinate system $(x,y,z)$,
where $\phi$ is the azimuthal angle measured on the disc surface
in the direction of flow, with $\phi = \pi/2$ at the ascending
node.\footnote{This definition of $\phi$ differs by $\pi/2$ from
that of \citet{pri96}.} The element of surface area is
\begin{equation}
  {\rm d}\vec{S}=\left[\vec{l}+(R\beta'\cos\phi+R\gamma'\sin\beta\sin\phi)\,
    \vec{e}_R\right]R\,{\rm d}R\,{\rm d}\phi,
\end{equation}
where the primes indicate differentiation with respect to $R$, and
$\vec{e}_R$ is the radial unit vector. The element of radiation flux ${\rm
d}F$ received from the ring source by unit area is
\begin{equation}
  {\rm d}F \propto \DF{{\rm d}\vec{S}\cdot\vec{p}}{|{\rm d}\vec{S}|}
    g_{\rm r}^3 I_{\rm r}(\nu_{\rm r}){\rm d}\Omega_{\rm r}
\end{equation}
where $I_{\rm r}(\nu_{\rm r})$ is the specific intensity measured by an
observer corotating with the ring source, $\vec{p}$ is the 3-momentum of
an incident photon from the source, $g_{\rm r}$ is a factor to describe the
shift of photon frequency along its path, which is equal to the ratio of 
the observed frequency from the illuminated disc to the emitted frequency 
from the ring source(see below). ${\rm d}\Omega_{\rm r}$ is the element of 
the solid angle subtended by the image of the ring source observed from 
the illuminated disc. The Balmer lines in the illuminated surface are 
formed as a result of photoionization by UV/X-ray radiation from a 
ring-like source. For simplicity, we assume that the Balmer line intensity 
is proportional to the radiation being intercepted by the disc 
\citep[see e.g. Fig.6a in][]{col89}, and thus the line emissivity 
$\varepsilon$ on the disc surface can take the form
$\varepsilon \propto \int{\rm d}F$. This is an approximate assumption. 
A detailed treatment of line emission requires solving the vertical 
structure as well as the radiative transfer in the disc explicitly, which 
is beyond the scope of this paper. We also assume that the line emission 
is isotropic in the comoving frame.

\subsection{The basic equations for a warped disc}
The dynamics of warping accretion discs have been discussed by a 
number of authors \citep{pap83,pap95,dem97,nay05}, and the static 
low-viscosity configurations were first calculated by \citet{iva97}. This 
section contains a short summary of the equations describing the structure
of the accretion disc models considered in this work. We follow the
formalism of \citet{pri92} but add a term to describe the
Lense-Thirring precession, to give
\begin{equation}
\begin{array}{lcl}
  \dot{\Sigma} & = &
    -\frac{1}{R} \Bracket{R \Sigma V_{\rm R}}'  \\
  \dot{\vec{L}} & = &
    -\frac{1}{R}\Bracket{R V_{\rm R} \vec{L}}' + \frac{1}{R}\vec{T_{\rm vis}}'
        + \vec{\Omega_{\rm P}}\times\vec{L}  \\
  \vec{T_{\rm vis}} &=&
    R^3 \Sigma \Bracket
    { \nu_1\Omega'\vec{l} + \frac{\nu_2}{2}\Omega\vec{l}' }
   \label{wdequt}
\end{array}
\end{equation}
\citep{che09}. Here we use a dot to denote $\partial/\partial t$,
and the prime symbol ($'$) to denote $\partial/\partial R$.
There are two viscosities, $\nu_1$ and $\nu_2$, where $\nu_1$
corresponds to the azimuthal shear, the standard shear viscosity
in a flat disc, and $\nu_2$ is the viscosity associated with vertical
shear motions, which smoothes out the warping. The Lense-Thirring
precession ${\bf\Omega}_{\rm p}$
is given by \citep[see e.g.][]{kum85} ${\bf\Omega}_{\rm p} =
\mbox{\boldmath$\omega$}_{\rm p}/R^3$, with:
\begin{equation}
\mbox{\boldmath$\omega$}_{\rm p}=\frac{2G{\bf J}}{c^2}~~~  {\rm and} ~~~
\vec{J} = acM\left(\frac{GM}{c^2}\right)\vec{j},
\end{equation}
where $\vec{J}=J\vec{j}$ is the angular momentum of the black hole.

We take both viscosities to have the same power law form, so that
\begin{equation}
\nu_1=\nu_{10}\left(\frac{R}{R_0}\right)^\eta ~~~ {\rm and} ~~~
\nu_2=\nu_{20}\left(\frac{R}{R_0}\right)^\eta,
\end{equation}
where $\nu_{10}$, $\nu_{20}$ and $\eta$ are all constants
and $R_0$ is some fixed radius. Although the selection of $R_0$ is in
principle arbitrary, it was chosen to be the Bardeen-Petterson radius
$R_{\rm BP}$($R_{\rm w}$), which is defined as $\vec{\omega_{\rm p}}/\nu_2$ 
in this work. For a steady-state warped disc,
the disc shape depends only on the radius $R_{\rm w}$. The exact value
of $R_{\rm w}$ is model-dependent and is subject to some uncertainties
\citep*{nat98,kin05,vol07}. To determine the value of $R_{\rm w}$, either 
the \citet{sha73} or the \citet{col90} model is used. Adopting the disc
properties derived by \citet{col90}, the warp radius in terms of the
Schwarzschild radius of the hole, $R_s$, is given by \citep{kin05}
\begin{eqnarray}
  \frac{R_{\rm w}}{R_{\rm s}} & =& 990 \left(\frac{\epsilon}{0.1}\right)^{1/4}
  \left(\frac{L}{0.1L_{\rm E}}\right)^{-1/4} M_8^{1/8} \times \left(\frac
  {\alpha_1}{0.03}\right)^{1/8} \nonumber \\ 
   & & \times \left(\frac{\alpha_2}{0.03}\right)^{-5/8}a^{5/8}.
\end{eqnarray}
Here, $\epsilon$ is the efficiency of the accretion process (i.e. $L =
\epsilon \dot{M} c^2$), $L$ is the accretion luminosity, $L_{\rm E}$ is the
Eddington limit, $M_8$ is the mass of the black hole in units of
$10^8$ M$_\odot$, and $a$ is the (dimensionless) spin of the black hole.

Assuming a Shakura-Sunyaev disc ({``}middle 
region"), the warp radius can be expressed as \citep{vol07}:
\begin{eqnarray}
  \frac{R_{\rm w}}{R_{\rm s}}= 3.6\times 10^3 a^{5/8}M_{8}^{1/8}\times 
   \left(\frac{\dot M c^2}{L_{\rm E}}\right)^{-1/4}\left(\frac{\nu_2}{\nu_1}
     \right)^{-5/8}\alpha_1^{-1/2}.
\label{eq:rw}
\end{eqnarray}
or, adopting the viscosity coefficient of a Shakura-Sunyaev disc as presented by 
\citet{fra02}, we have
\begin{eqnarray}
\frac{R_{\rm w}}{R_{\rm s}} &=& 1.41\times 10^4 a^{4/7}\left(\frac{\nu_2}{\nu_1}
     \right)^{-4/7} \left(\frac{\alpha_1}{0.03}\right)^{-16/35}
     \left(\frac{\dot M c^2}{L_{\rm E}}\right)^{-6/35} \nonumber \\
     & & \times M_8^{4/35}
\end{eqnarray}
For the typically expected standard values of AGNs, $R_{\rm w}$ is hundreds to 
thousands of $R_{\rm s}$.

\subsection{Photon motions in the background metric}
\label{equat}
We review properties of the Kerr metric and formulae
for its particle orbits, and summarize here the basic equations
relevant to this paper. In Boyer-Lindquist coordinates,
the Kerr metric is given by \citep{cha83}:
\begin{eqnarray}
{\rm d}s^{2} & = & -{\rm e}^{\rm 2\nu}{\rm d}t^{2}+{\rm e}^{2\psi}
   ({\rm d}\phi-\omega{\rm d}t)^2+\frac{\Sigma}{\Delta}{\rm d}r^{2} + 
    \Sigma {\rm d}\theta^{2} ,
\end{eqnarray}
where
\[{\rm e}^{2\nu}=\Sigma\Delta/A,\,{\rm e}^{\rm 2\psi} = \sin^{2}\theta 
   A/\Sigma,\,\omega=2Mar/A,\]
\[\Sigma=r^{2}+a^{2}\cos^{2}\theta,\,\Delta=r^{2}+a^{2}-2Mr,\]
\[A=(r^{2}+a^{2})^{2}-a^{2}\Delta\sin^{2}\theta.\]
Here $M$, $a$ are the black hole mass and specific angular momentum,
respectively. The equation of motion governing the orbital trajectory
in the $r\theta$-plane is \citep{bar72}
\begin{eqnarray}
    \int_{r_{\rm e}}^r \frac{{\rm d}r}{\sqrt{R(r)}} = \pm
        \int_{\theta_{\rm e}}^\theta \frac{{\rm d}\theta}{\sqrt{\Theta(
       \theta)}} \;,
    \label{r-thet}
\end{eqnarray}
where $r_{\rm e}$ and $\theta_{\rm e}$ are the starting values of $r$ and
$\theta$. The $\phi$-coordinate along the trajectory is calculated by
\citep{wil72,vie93}
\begin{equation}
\int_{\phi_{\rm e}}^{\phi}{\rm d}\phi =
\int_{\theta_{\rm e}}^\theta\frac{\lambda{\rm d}\theta}{\sin^2\theta
  \sqrt{\Theta(\theta)}} + \int_{r_{\rm e}}^r \frac{(2ar-\lambda a^2){\rm d}r}
  {\Delta\sqrt{R(r)}}.
\label{phi-thet}
\end{equation}

For photons emitted from the broad line region with radii $>$100$r_g$
propagating to infinity, we can neglect terms of order $1/r^2$ and 
higher in equations~(\ref{r-thet}) and (\ref{phi-thet}), the integral 
over $\theta$ can be worked out in terms of a trigonometric integral
\begin{eqnarray}
  \int_{\pi/2}^\vartheta \frac{{\rm d}\vartheta}{\sqrt{\Theta(\vartheta)}}
         &=&\frac{1}{\sqrt{\lambda^2 + q^2}}\, \rm {sin^{-1}
       \left(\mu/\mu_+\right)},
    \label{mu_int}
\end{eqnarray}
\begin{eqnarray}
  \int_{\pi/2}^{\vartheta}\frac{\lambda{\rm d}\vartheta}{{\rm
      sin}^{2}\vartheta \sqrt{\Theta(\vartheta)}}
 &=& \pm \rm{sin^{-1}\sqrt{\frac{(1-\mu_+^{2})\mu^{2}}{\mu_+^{2}(1-\mu^{2})}}}.
    \label{phi_int}
\end{eqnarray}
Where $\mu = \cos\theta$ and $0\leq \mu < \mu_+$, $\mu_+ = 
\sqrt{q^2/(\lambda^2 + q^2)}$. The integral over $r$ can be 
worked out using inverse Jacobian elliptic integrals 
\citep*[see e.g.][]{cad98,wu07}.

The specific flux density $F_{\rm o}(\nu_{\rm o})$ at frequency
$\nu_{\rm o}$ is given by \citep{cun75}
\begin{eqnarray}
 F_{\rm o}(\nu_{\rm o})
  & = & \int g^3 I_{\rm e}(\nu_{\rm e}) {\rm d}\Omega_{\rm obs} \nonumber \\
  & = & \frac{q}{r_{\rm o}^2\beta\sin\vartheta_{\rm o}}
     \int \varepsilon g^4 \delta(\nu_{\rm o}-g\nu_{\rm e})
     \frac{\partial(\lambda,q)}{\partial(r,g)}\;{\rm d}r\;{\rm d}g.
    \label{feo}
\end{eqnarray}
where ${\rm d}\Omega_{\rm obs}$ is the element of the solid angle
subtended by the image of the disc on the observer's sky and $g$ is 
a factor to describe the shift of photon frequency along its path 
and is equal to the ratio of the observed to the emitted frequency. the 
emissivity $\varepsilon$ in the integrand is calculated by
\begin{eqnarray}
 \varepsilon
  & = & \int\DF{{\rm d}\vec{S}\cdot\vec{p}}{|{\rm d}\vec{S}|}
    g_{\rm r}^3 I_{\rm r}(\nu_{\rm r}){\rm d}\Omega_{\rm r} 
    \label{epsilon}
\end{eqnarray}
where $g_{\rm r}$ is the redshift for radiation between the emitter's and the 
observer's frame in the special case of two orbiting systems, and is 
given by \citep{vie93}
\begin{eqnarray}
    g_{\rm r} & = & \frac{\gamma_{\rm d} {\rm e_d}^\nu(1 - \Omega_{\rm d}\lambda)}
         {\gamma_{\rm r} {\rm e_r}^\nu(1 - \Omega_r\lambda)},
     \label{gvalue}
\end{eqnarray}
where $\gamma_{\rm d} and \gamma_{\rm r}$ are the Lorentz factor measured 
in the locally non-ratating frame(LNRF). The element of solid angle 
${\rm d}\Omega_{\rm r}$ is
\begin{eqnarray}
    {\rm d}\Omega_{\rm r} = \frac{{\rm d}\alpha {\rm d}\beta}{r_{\rm d}^2}
    = \frac{1}{r_{\rm d}^2}\frac{\partial(\alpha,\beta)}{\partial(\lambda,q)}
    \frac{\partial(\lambda,q)}{\partial(r,g_r)}\;{\rm d}r\;{\rm d}g_r,
    \label{solid_obs}
\end{eqnarray}
where $r_{\rm d}$ is the distance from the illuminated point on the disc 
to the black hole. The two impact parameters $\alpha$ and $\beta$, first
introduced by \citet{cun73}, are defined as
\begin{eqnarray}
    \alpha = - \frac{rp^{(\varphi)}}{p^{(t)}}~~~{\rm and}~~~
    \beta = \frac{rp^{(\theta)}}{p^{(t)}}\!, \label{alp_beta}
\end{eqnarray}
where $p^{(t)}, p^{(\theta)}, p^{(\varphi)}$ are the tetrad(or LNRF) 
components of the four-momentum \citep[see e.g.][]{pin77,cha83}.

\subsection{Method of calculation }
\label{meth}

We now turn to how to calculate the line profiles numerically. A 
steady-state warped disc structure is first obtained by solving
the dynamical equation numerically by setting $\partial\vec{L}/ 
\partial t=0$ in equation~(\ref{wdequt}). The disc is divided 
into a number of arbitrarily narrow rings, each such emitting ring 
being denoted by its radius $r_{\rm i}$ and weights $\omega_{\rm i}$, 
provided by an algorithm developed by Rybicki G. B. \citep{pre92} 
for Gauss-Legendre integration. The main numerical 
procedures for computing the line profiles are as follows:
\begin{enumerate}
\item Using the numerical disc structure obtained first, the values 
of two Eulerian angles $\beta(R,t)$ and $\gamma(R,t)$ at an arbitrary 
point are calculated by means of cubic spline interpolation.

\item The relevant disc system parameters are specified: $R_{\rm in}, R_{\rm
out},R_{\rm r}, R_{\rm w}$, $i$, $\eta$ and $\nu_2/\nu_1$.

\item For a given couple~($r_{\rm i},g$) of a ring, the two constants of
motion $\lambda$ and $q$ are determined if they exist.

\item For each g, the integration over $r$ of equation~(\ref{feo})
can be replaced by 
\begin{eqnarray}
  F_{\rm o}(\nu_{\rm o}) &=& \sum_{i=1}^n \frac{q \varepsilon \nu_{\rm o}^4}
  {r_{\rm o}^2 \nu_{\rm e}^4 \beta\sin\vartheta_{\rm o}}\left.\frac
  {\partial(\lambda,q)}{\partial(r,g)}\right|_{\rm r=r_i}\omega_{\rm i}.
  \label{feo4}
\end{eqnarray}
where the emissivity $\varepsilon$ in equation~(\ref{feo4}) is 
calculated numerically by equation~(\ref{epsilon}).
\end{enumerate}

From the above formula, one can determine the line flux from the disc 
at an arbitrary frequency $\nu_{\rm o}$. The observed line profile as a
function of frequency $\nu_{\rm o}$ is finally obtained in this way.

\section{Results}
\label{result}

In the accretion disc model, the double-peaked emission lines are
radiated from the disc region between around several hundred
gravitational radii to more than 2000$r_{\rm g}$; here, $r_{\rm g}$ is 
the gravitational radius, and the widths of the double-peaked lines 
range from several thousand to nearly 40,000~$\rm km\,s^{-1}$
\citep{wang05}. In our model, all of the parameters of the warped
disc are set to be free. The warp radius $R_{\rm w}$ is a key parameter 
for describing the disc shape. To determine the value of $R_{\rm w}$, 
either the \citet{sha73} or the \citet{col90} model is used. As far as 
the outer region of the disc is concerned, the self-gravity of the disc 
must be taken into account at large radii. Beyond a certain radius, the 
disc becomes gravitationally unstable and can be disrupted. The typical 
range of unstable region is in the order of $10^3$ to a few $10^4$ 
Schwarzschild radius \citep{col99}. Thus, the reasonable range of the 
disc is from 100$r_g$ to 2000$r_g$. In our calculation, the frequency 
ranges from 4.32 to 4.78 in units of $10^{14}Hz$, and 180 bins are used. 
Considering the broadening arising from electron scattering or turbulence, 
all our results are smoothed by convolution with a 3$\sigma$ Gaussian.

\subsection{The numerical steady-state warped disc solutions}

\begin{figure}
\includegraphics[width=1.0\hsize]{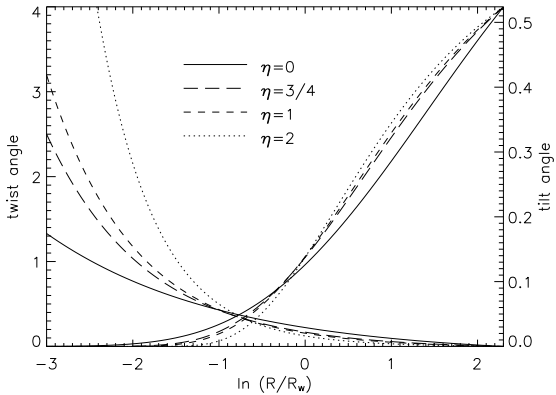}
 \caption{The distribution of $\beta$ and $\gamma$ as a function of radius.
The twist angle $\gamma$ is measured in units of 2$\pi$, and the tilt angle 
$\beta$ is shown in radians. The solid, long-dashed, short-dashed and dotted 
lines show the results for $\eta=0, 3/4, 1$ and 2, respectively.
 \label{wpdisc}}
\end{figure}

We numerically solved equation~(\ref{wdequt}), setting all the time 
differentials to zero, in order to find a steady-state disc structure.
As a boundary condition, we set $\beta=\pi/6$ and $\gamma=0$ 
at the outer boundary. The calculation region is $-4.6<x<2.3$, 
where $x=\ln(R/R_{\rm w})$, or, equivalently, $R_{in}=R_{\rm w}{\rm e}^{-4.6}
\approx0.01R_{\rm w}$, and $R_{\rm out}=R_{\rm w}{\rm e}^{2.3}\approx10R_{\rm w}$.
The calculated distributions of $\beta$ and $\gamma$ as a function of radius 
are plotted in Fig.~\ref{wpdisc}.

\subsection{Line profiles from twisted warping discs}

\begin{figure}
\includegraphics[width=1.0\hsize]{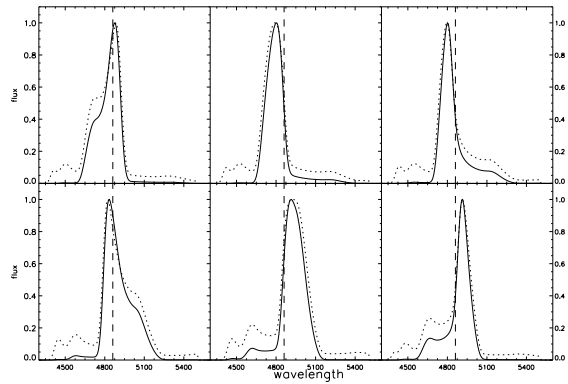}
  \caption{Comparison of the H$\beta$ line profiles between a simplified model 
with no relativistic effects(solid line) and one with the ralativistic 
treatment(dotted line). The parameters of two Eulerian angles and the disc 
range are identical to those presented by \citet{bac99}. The azimuthal viewing 
angle along the XY-plane varies from $0^{\circ}$ to $300^{\circ}$ in steps 
of $60^{\circ}$ from the top left-hand panel to the bottom right-hand panel. 
The vertical dashed line corresponds to the wavelength position of H$\beta$ 
in the rest frame.
 \label{comb}}
\end{figure}

We use the numerical code developed by \citet{wu08} to evaluate the line 
profiles for the numerical disc structure obtained. We first calculated  
the line profiles of a simplified model with no relativistic effects included 
in equations of propagation of photons and compared them with the relativistic 
treatment. The parameters of the two Eulerian angles and the disc range are 
identical to those found by \citet{bac99}, but we used a ring-like primary 
source and a linear response for the reprocessing to the ionizing flux. The 
primary source radius is set to $R_r=1.5r_{\rm g}$. We reproduced the results 
obtained by \citet{bac99} on a qualitative level. A comparison between the 
H$\beta$ line profiles for a simplified model with no relativistic 
effects(solid line) and those with the ralativistic treatment(dotted line) 
is show in Fig.~\ref{comb}. The influence of the gravitational lensing, 
which concentrates the radiation from the innermost region towards the 
equatorial plane on the line profle, is significant. 

\begin{figure}
\includegraphics[width=1.0\hsize]{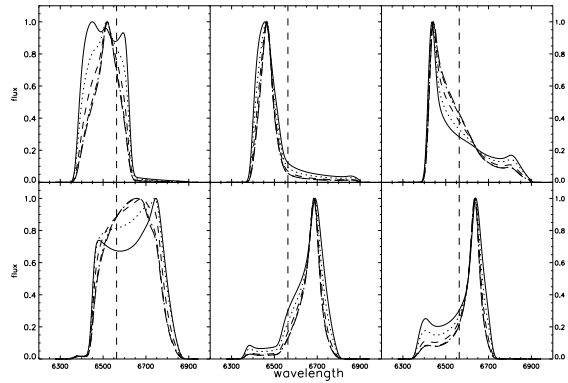}
  \caption{Comparison of the H$\alpha$ line profiles with different 
primary source radii $R_{\rm s}$: $R_{\rm s}=1.5r_{\rm g}$(solid line), 
$2.0r_{\rm g}$(dotted line), $3.0r_{\rm g}$(short dashed line), 
$6.0r_{\rm g}$(dot-dashed line), $10.0r_{\rm g}$(long dashed line). The 
other parameters are $R_{\rm in} = 150r_{\rm g}$, $R_{\rm out}= 1200 r_{\rm g}$, 
$R_{\rm w}=600r_{\rm g}$, $\eta=1$, $\nu_2/\nu_1=10$ and $i=30^{\circ}$. The 
azimuthal viewing angle along the XY-plane varies from $0^{\circ}$ to 
$300^{\circ}$ in steps of $60^{\circ}$ from top left-hand panel to bottom 
right-hand panel. The vertical dashed line corresponds to the wavelength 
position of H$\alpha$ in the rest frame.
 \label{sourcer}}
\end{figure} 

The dependence of the H$\alpha$ line profiles on the primary source radius 
$R_{\rm r}$ is ploted in Fig.~\ref{sourcer}. The radii with 
respect to the different lines are: $R_{\rm r}=1.5r_{\rm g}$(solid line), 
$2.0r_{\rm g}$(dotted line), $3.0r_{\rm g}$(short-dashed line), 
$6.0r_{\rm g}$(dot-dashed line), $10.0r_{\rm g}$(long-dashed line). The 
other parameters are $R_{\rm in} = 150r_{\rm g}$, $R_{\rm out}= 1200r_{\rm g}$, 
$R_{\rm w}=600r_{\rm g}$, $\eta=1$, $\nu_2/\nu_1=10$ and $i=30^{\circ}$. 
The azimuthal viewing angle along the XY-plane varies from $0^{\circ}$ to 
$300^{\circ}$ in steps of $60^{\circ}$ from the top left-hand to the bottom 
right-hand panel. The vertical dashed line corresponds to the wavelength 
position of H$\alpha$ in the rest frame. From this figure, it can be seen 
that the line profiles are not sensitive to a change in the radius of the 
primary source in the range from $3.0r_{\rm g}$ to $10.0r_{\rm g}$, but that 
they have a significant dependence when it approaches the innermost stable orbit.

\begin{figure}
\includegraphics[width=1.0\hsize]{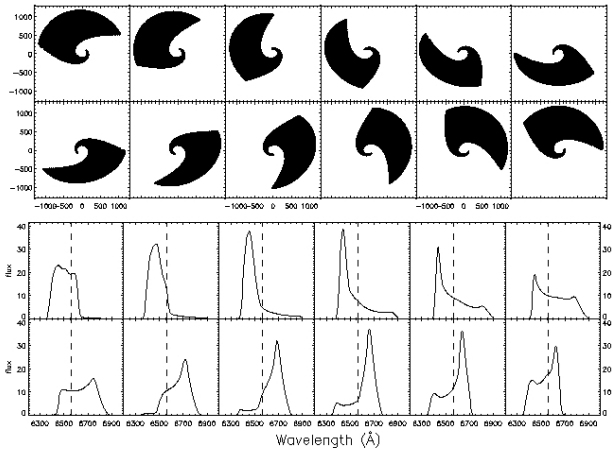}
  \caption{Images~(upper panels) of the illuminated area of the disc and 
the H$\alpha$ line profiles~(lower panels) computed by our code for the 
twisted warped disc case for $i=30^{\circ}$; the disc zone is from $R_{\rm in} 
= 150r_{\rm g}$ to $R_{\rm out}= 1200 r_{\rm g}$. The Bardeen-Petterson 
radius is set to $R_{\rm w}=600r_{\rm g}$, $\eta=3/4$ and $\nu_2/\nu_1=10$. 
The azimuthal viewing angle along the XY-plane varies from $0^{\circ}$ to 
$330^{\circ}$ in steps of $30^{\circ}$ from the top left-hand panel to the 
bottom right-hand panel. 
 \label{twist1}}
\end{figure}

The images of the illuminated area and the H$\alpha$ line profiles 
computed by our code for a steady-state twisted warped disc are shown 
in Fig.~\ref{twist1}. The disc zone is from $R_{\rm in}=150r_{\rm g}$
to $R_{\rm out}= 1200 r_{\rm g}$, the warp radius is set to $R_{\rm w}
=600r_{\rm g}$, and the other parameters are $i=30^{\circ}, \eta=3/4$ 
and $\nu_2/\nu_1=10$. The azimuthal viewing angle along the XY-plane 
varies from $0^{\circ}$ to $330^{\circ}$ in steps of $30^{\circ}$ 
from the top left-hand to the bottom right-hand panel. The image contains 
$400\times 360$ pixels. From the images, we can see that, moving inwards 
from the outer radius of the disc, the warp is twisted by an angle of 
$\sim\pi$ before being flattened efficiently into the aligned plane, as 
is also shown in Fig.~\ref{wpdisc}. 

\begin{figure}
\includegraphics[width=1.0\hsize]{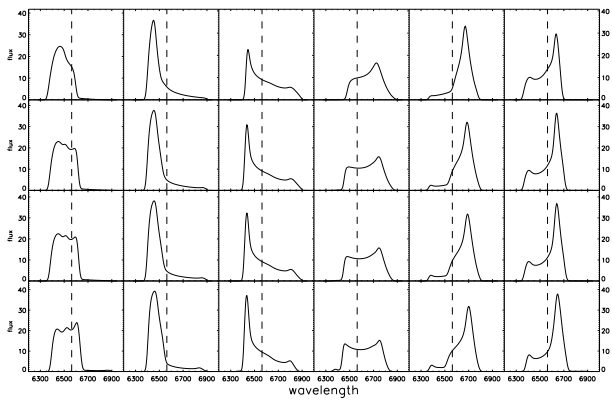}
  \caption{Comparison of the H$\alpha$ line profiles for a warped disc 
with various values of the power-law index of the viscosity: 
$\eta = 0,\,3/4,\,1,\,2$~(from top to bottom rows). 
The other parameters are $i=30^{\circ},\; R_{\rm in}=150r_{\rm g}$, 
$R_{\rm out}= 1200 r_{\rm g}$ and $R_{\rm w}= 600 r_{\rm g}$. The azimuthal 
viewing angle along the XY-plane varies from $0^{\circ}$ to $300^{\circ}$ 
in steps of $60^{\circ}$~(from the left- to the right-hand side).
 \label{powerl}}
\end{figure}

\begin{figure}
\includegraphics[width=1.0\hsize]{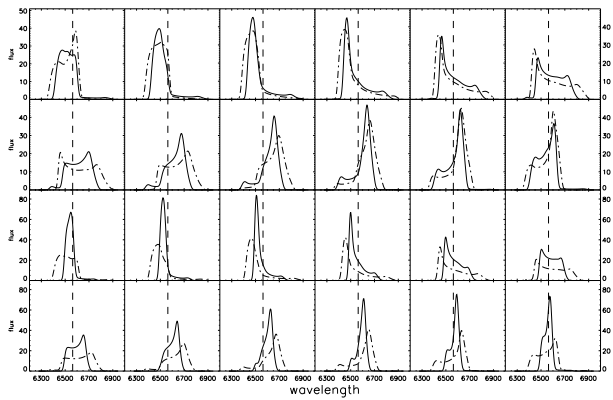}
  \caption{Comparison of the H$\alpha$ line profiles with various values 
of the Bardeen-Petterson radius or viewing angle. 
Upper two rows: the parameters are $i=30^{\circ},\; 
R_{\rm in}=150r_{\rm g}, R_{\rm out}= 2000 r_{\rm g}$, $\eta=3/4$, 
$\nu_2/\nu_1=10$ and $R_{\rm w}=1000r_{\rm g}$~(solid line), 
$R_{\rm w}=500r_{\rm g}$~(dot-dashed line).
Lower two rows: the parameters are $R_{\rm in} = 150r_{\rm g}$, 
$R_{\rm out}= 1500 r_{\rm g}$, $R_{\rm w}=800r_{\rm g}$, $\eta=1$, 
$\nu_2/\nu_1=10$ and $i=15^{\circ}$~(solid line), $i=30^{\circ}$~
(dot-dashed line). The azimuthal angle along the XY-plane varies 
from $0^{\circ}$ to $330^{\circ}$ in steps of $30^{\circ}$ from 
the top left-hand to the bottom right-hand panel. 
 \label{bprang}}
\end{figure}

The influence of the power-law index of the viscosity on the line 
profiles for $\eta = 0,\,3/4,\,1,\,2$~(from top to bottom panels) is 
shown in Fig.~\ref{powerl}. The disc zone is from $R_{\rm in} = 
150r_{\rm g}$ to $R_{\rm out}=1200 r_{\rm g}$; the other parameters 
are $i=30^{\circ}$, $R_{\rm w}= 600 r_{\rm g}$.The azimuthal viewing 
angle along the XY-plane varies from $0^{\circ}$ to $300^{\circ}$ in 
steps of $60^{\circ}$~(from the left- to the right-hand side). 
The general behaviour of the line profile is similar. 
Fig.~\ref{bprang} shows the dependence of the profiles on the warp 
radius and viewing angle. The upper panels are for $i=30^{\circ},\; 
R_{\rm in}=150r_{\rm g}, R_{\rm out}= 2000 r_{\rm g}$, $\eta=3/4$,
$\nu_2/\nu_1=10$ and $R_{\rm w}=1000r_{\rm g}$~(solid line),
$R_{\rm w}=500r_{\rm g}$~(dot-dashed line). The lower panels 
show the profile dependence on the viewing angles for $i=15^{\circ}$~
(solid line) and $i=30^{\circ}$~(dot-dashed line). The other parameters 
are $R_{\rm in} = 150r_{\rm g}$, $R_{\rm out}= 1500 r_{\rm g}$, 
$R_{\rm w}=800r_{\rm g}$, $\eta=1$ and $\nu_2/\nu_1=10$. 

Generally speaking, the profiles of the line emission from the warped 
disc induced by the Bardeen-Petterson effect are nonsymmetrical and red- 
or blue-frequency-shifted in most cases for which the primary source radius 
is large. When the primary source radius approaches the innermost stable 
orbit, double-peaked or triplet-peaked line profiles are present in most 
cases. Thus, a warped disc induced by the Bardeen-Petterson effect provides 
a mechanism for producing the double- or triplet-peaked line profiles that 
were observed in the Sloan Digital Sky Survey. The triplet-peaked line 
profile, such as for SDSS J084205.57+075925.5 and SDSS J232721.96+152437.3 
\citep[see e.g.][]{wu08}, may be a {``}signature" of a warped disc. Future 
monitoring of line profile variability in these triplet sources will  
provide a critical test of the warped disc model.

\section{Summary}

\label{sum} We computed the Balmer emission-line profiles that result from
the reprocessing of emission using a numerical model of a warped disc 
induced by the Bardeen-Petterson effect, including all relativistic 
effects for radiation propagation, and subject to various shadowing 
effects associated with disc warping. We numerically solved the disc 
structure for a steady-state disc shape. To simplify the analysis, we 
took both viscosities to have the same power-law form, and thus the ratio 
$\nu_2/\nu_1$ is independent of radius. For simplicity, we assumed 
that the disc is illuminated by a ring-like central source, which is a 
rough approximation, that the line emissivity is proportional to the 
continuum light intercepted by the accretion disc, and that line emission 
is isotropic. Our conclusions are as follows.
\begin{enumerate}
\item From the numerical results, we can see that, moving inwards 
from the outer radius of the disc, the warp is twisted by an angle of 
$\sim\pi$ before being flattened efficiently into the equatorial plane. 
\item For a given warped disc, there are two angles that determine the 
observed line profile, the inclination angle and the azimuthal viewing 
angle resulting from the non-axisymmetry of the disc warping.
\item For less twisted warped disc induced by the Bardeen-Petterson effect, 
the asymmetrical and frequency-shifted single-peaked line profiles are
produced in most cases when the primary source radius is large(from 3$r_g$  
to 10$r_g$). The double- or triplet-peaked line profiles presented in most 
cases occur when the primary source radius approaches the marginally stable 
orbit of an extreme Kerr black hole.

\end{enumerate}

The line profiles depend strongly on the twisting structure of the disc 
\citep{wu08} as well as on the radius of the illuminating source. The 
Doppler effect and gravitational focusing concentrate the radiation from the 
innermost region towards the equatorial plane, which increases the returning 
flux \citep{cun75,cun76}. This may be the reason why the illuminating source 
radius has a strong influence on the line profile. Knowledge of the 
twisting structure may allow the determination of some important 
characteristics of AGNs, such as the spin of the black hole and the 
viscosities of the disc. We assume that the ratio $\nu_2/\nu_1$ is independent 
of radius in the calculations, but the relation between $\nu_1$ and $\nu_2$ 
is very uncertain, and may influence the disc twisting structure. It is 
also important to emphasize that, although we have 
analyzed the Bardeen-Petterson effect in this work, the four mechanisms 
mentioned above are not mutually exclusive. This means that more than one 
mechanism might be operating in a given system.

\section{acknowledgments}
We would like to thank the anonymous referee for helpful
suggestions and comments, which improved and clarified our paper.
This work was supported in part by the Natural Science Foundation of China
(grants 10773024, 10833002, 10821302, and 10825314),
Bairen Program of Chinese Academy of Sciences,
and the National Basic Research Program of China (973 Program 2009CB824800).

\label{lastpage}
\end{document}